\documentclass[onecolumn,floatfix,superscriptaddress,showpacs,showkeys,nofootinbib,preprint]{revtex4}

\usepackage{epsfig,hhline}
\usepackage{amssymb,latexsym,amsmath}
\newcommand{\eq}[1]{\begin{align} #1 \end{align}}

\begin{document}
\title{Pion Number Fluctuations and Correlations \\in the Statistical System
with Fixed Isospin}

 \author{V. V. Begun}
 \affiliation{Bogolyubov Institute for Theoretical Physics,
 Kiev, Ukraine}
 \affiliation{Frankfurt Institute for Advanced Studies, Frankfurt,
 Germany}

 \author{M. I. Gorenstein}
 \affiliation{Bogolyubov Institute for Theoretical Physics,
 Kiev, Ukraine}
 \affiliation{Frankfurt Institute for Advanced Studies, Frankfurt, Germany}

 \author{O. A. Mogilevsky}
 \affiliation{Bogolyubov Institute for Theoretical Physics,
 Kiev, Ukraine}

\begin{abstract}
The statistical system of pions with zero total isospin is
studied.  The suppression effects for the average yields due to
isospin conservation are the same for $\pi^0$, $\pi^+$ and
$\pi^-$. However, a behavior of the corresponding particle number
fluctuations are different. For neutral pions there is the
enhancement of the fluctuations, whereas for charged pions the
isospin conservation suppresses fluctuations.  The correlations
between the numbers of charged and neutral pions are observed for
finite systems.  This causes a maximum of the total pion number
fluctuations for small systems. The thermodynamic limit values for
the scaled variances of neutral and charged pions are calculated.
The enhancements of the fluctuations due to Bose statistics are
found and discussed.

\end{abstract}



\maketitle

%
%
%
\section{Introduction}

The aim of the present paper is to study some aspects of
non-Abelian symmetries in the statistical models. We consider
SU(2)-isospin symmetry group for the pion system. The role of the
isospin conservation in a many body system was first considered in
the pioneering paper of Bethe \cite{Bethe}. Many efforts were then
aimed at studies of the pion system with fixed isospin
\cite{Pais,Kretz}.  An effective theoretical formalism for
non-Abelian symmetries in the  statistical mechanics was developed
in Ref.~\cite{Turko} on the basis of the group projection
technique.  It allowed to consider the impact of the isospin
conservation on the particle abundances and the form of their
momentum spectra in the statistical models of hadron production
\cite{Muller,Blumel}.   The group projection technique was also
used to calculate the colorless partition function of the
quark-gluon gas with SU(N$_c$)-color symmetry  \cite{Elze,GM}.

Our primary interest in the present paper is to study an influence
of non-Abelian charge conservation on the particle number
fluctuations.  It was recently found \cite{CE} that exact
conservation of Abelian (additive) charges causes  the suppression
of the particle number fluctuations.
In the present study we restrict our consideration to the simplest
statistical system with non-Abelian symmetry -- an ideal pion gas
with
zero isospin $I=0$.  Most discussions are done within Boltzmann
statistics. 
This makes possible to obtain transparent analytical results and
compare them with those in the canonical ensemble for zero
electric charge $Q=0$.

The paper is organized as the following. In Section~\ref{sec-Z} we
consider the partition function and total number of pions (the
mean value and scaled variance).  In Section~\ref{sec-pions} we
calculate the mean multiplicities, fluctuations and correlations
for $\pi^+$, $\pi^-$ and $\pi^0$ mesons. In Section~\ref{sec-Bose}
we consider the specific effects due  to Bose statistics. The
Section~\ref{sec-summary} summarizes the paper.
%
%
\section{Partition Function and total number of pions}\label{sec-Z}
%
%
The partition function of the ideal Boltzmann gas of pions
$\pi^+,~\pi^-,~\pi^0$ in the grand canonical ensemble (GCE) reads,
\eq{\label{gce}
Z_{gce}~=~\sum_{N_0,N_+,N_-=0}^{\infty}\frac{\left(\lambda_0~z\right)^{N_0}}{N_0!}~
\frac{\left(\lambda_+~z\right)^{N_+}}{N_+!}
~\frac{\left(\lambda_-~z\right)^{N_-}}{N_-!}~=~\exp\left[\left(\lambda_0+\lambda_+
+\lambda_-\right)z\right]~,
}
where $z$ is the one-particle partition function,
 \eq{\label{z}
 z \;=\; \frac{V}{2\pi^2}\,\int_0^{\infty}p^2dp\;
           \exp\left(-\,\frac{\sqrt{p^2+m^2}}{T}\right)
     \;=\; \frac{V}{2\pi^2}~Tm^2~K_2\left(\frac{m}{T}\right)\;.
 }
Here $V$ and $T$ are the system volume and temperature, $m$ is the
pion mass (we neglect a small difference between the masses of
charged and neutral pions), and $\;K_2\;$ is the modified Hankel
function.  The auxiliary parameters $\lambda_j$ with $j=0,~+,~-$
are introduced  to calculate the mean pion multiplicities,
fluctuations and correlations.  We take
$\lambda_j\equiv 1$ in the final formulae.

In the case of exact charge conservation, i.e. in the canonical
ensemble (CE) with zero charge $Q=0$, the partition function is
(see, e.g. \cite{RD,CE}):
 \eq{\label{Zce}
 Z_{Q=0} &\;=\;
 \sum_{N_0,N_+,N_-=0}^{\infty}\delta\left(N_+~-~N_-\right)~
 \frac{\left(\lambda_0~z\right)^{N_0}}{N_0!}~
\frac{\left(\lambda_+~z\right)^{N_+}}{N_+!}
~\frac{\left(\lambda_-~z\right)^{N_-}}{N_-!}\nonumber \\
 &=~\exp\left(\lambda_0~z\right)~\frac{1}{2\pi}\int_0^{2\pi}d\phi\;
 \exp\left[~ z~
     \left(\lambda_+\exp[i\phi] \;+\; \lambda_-\exp[-i\phi]\right)\;\right]\;.
 }
Note that an exact charge conservation in the CE (\ref{Zce}) does
not affect the neutral pions. Their number distribution remains
the Poissonian one, the same as in the GCE (\ref{gce}).

The partition function  with total isospin $I=0$ can be obtained
using group projection technique. Pions are transformed under
vector  (adjoint) representation of the SU(2) group. This
group has three parameters which can be chosen as Euler angles
$\vec{\alpha}=\alpha,\beta,\gamma$. In this case the diagonal
matrix elements have the following form \cite{Wigner}:
 \eq{\label{D}
  D_{\pm 1,\pm 1}^1(\alpha,\beta,\gamma)
 \;=\; e^{\pm i(\alpha + \gamma)}\left(\frac{1+\cos(\beta)}{2}\right)\;,\qquad
 D_{0,0}^1(\alpha,\beta,\gamma)
 \;=\; \cos(\beta)\;.
 }
 The partition function is then presented as
\cite{Blumel,Turko2}:
 \eq{\label{Z0-Group}
 Z_{I=0}~
 & =\; \int d\mu\;  \sum_{N_0,N_+,N_-=0}^{\infty}
\frac{\left[ \lambda_0~z~D_{0,0}^{1}(\vec{\alpha})\right]^{N_0}}{N_0!}~
\frac{\left[\lambda_+~z~D_{1,1}^{1}(\vec{\alpha})\right]^{N_+}}{N_+!}
       ~\frac{\left[\lambda_-~z~D_{-1,-1}^{1}(\vec{\alpha})\right]^{N_-}}{N_-!} \nonumber \\
 & =\; \int d\mu\;\exp\left[
       \lambda_0~z~D_{0,0}^{1}(\vec{\alpha})~+~\lambda_+~z~D_{1,1}^{1}(\vec{\alpha})
       ~+~\lambda_-~z~D_{-1,-1}^{1}(\vec{\alpha})\right]\;.
}
%
%
%
%
Substituting explicit expressions for the Haar group measure
$d\mu$ and matrix elements $D_{t_3,t_3}^{t}$ (\ref{D})  in
Eq.~(\ref{Z0-Group}), one obtains:
%
%
%
 \eq{\label{Z-Boltz}
 &Z_{I=0} \;=\;
 \frac{1}{8\pi^2}\int_0^{2\pi}d\alpha\int_0^{2\pi}d\gamma\int_0^{\pi}d\beta
    \sin\beta
 \\
 &\times\;
    \exp\left[\lambda_0\,z\,\cos\beta
    \;+\; z\left(\frac{1+\cos\beta}{2}\right)~
    \bigg(\lambda_+~\exp[i(\alpha+\gamma)]
    \;+\; \lambda_-~\exp[-i(\alpha+\gamma)]\bigg)\right]\;.
 \nonumber }
The change of variables $\phi=\alpha+\gamma$,
$\varphi=(\alpha-\gamma)/2$, $\cos(\beta)=x$ and integration over
$\varphi$ gives:
 \eq{\label{ZI=0}
 Z_{I=0} &\;=\;
 \frac{1}{4\pi}\int_0^{2\pi}d\phi\int_{-1}^{1}dx
 \exp\left[\lambda_0\,z\,x \;+\; z\,\frac{1+x}{2}
     \left(\lambda_+\exp[i\phi] \;+\; \lambda_-\exp[-i\phi]\right)\right]\;.
 }
Comparing $Z_{I=0}$ (\ref{ZI=0}) with the partition function
$Z_{Q=0}$ (\ref{Zce}) one observes an additional $x$-integration
in Eq.~(\ref{ZI=0}). It reflects a presence of the particle number
correlations between neutral and charged pions which were absent
in the GCE and CE.

\begin{figure}[ht!]
\epsfig{file=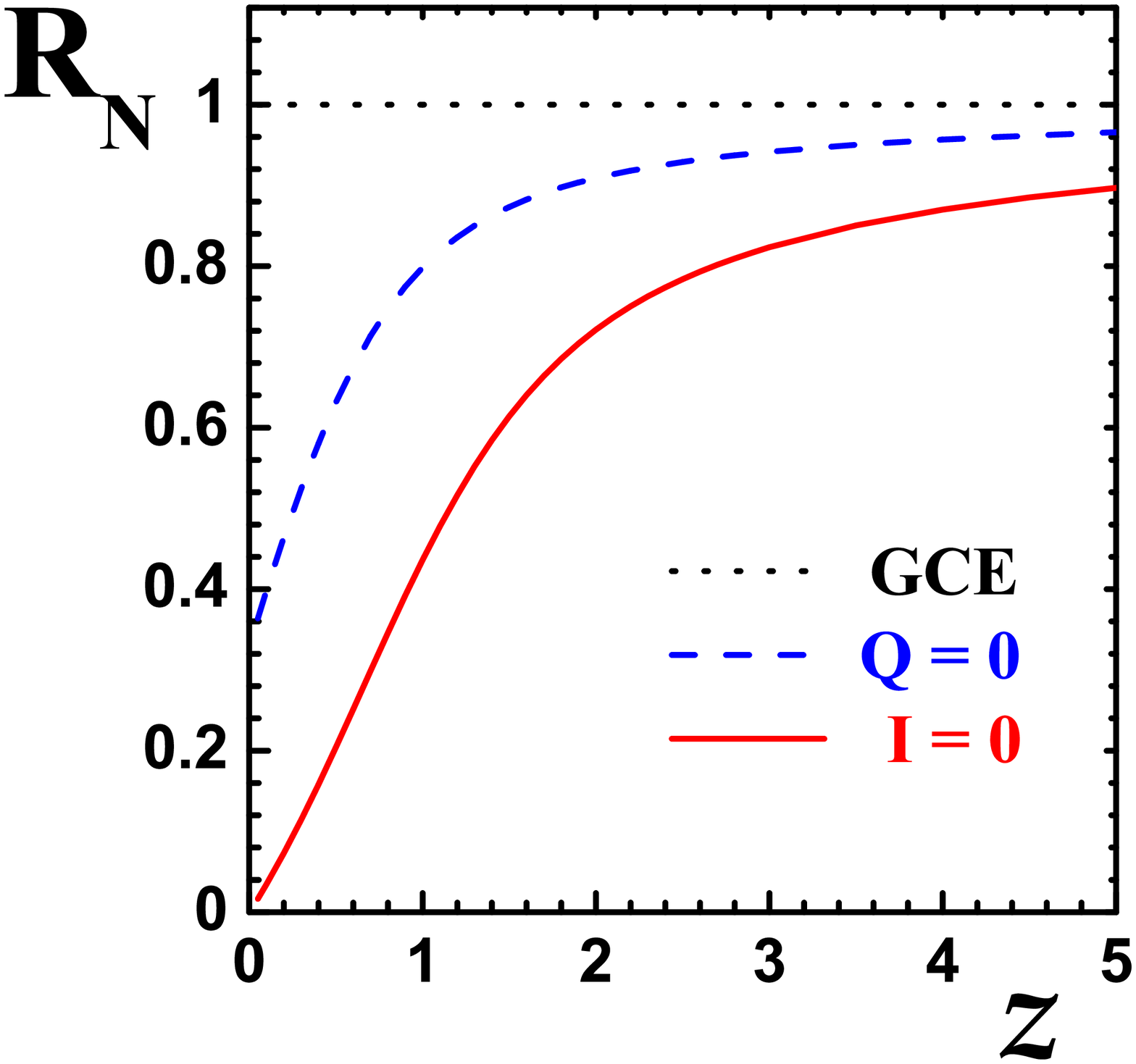,width=0.48\textwidth}~~
\epsfig{file=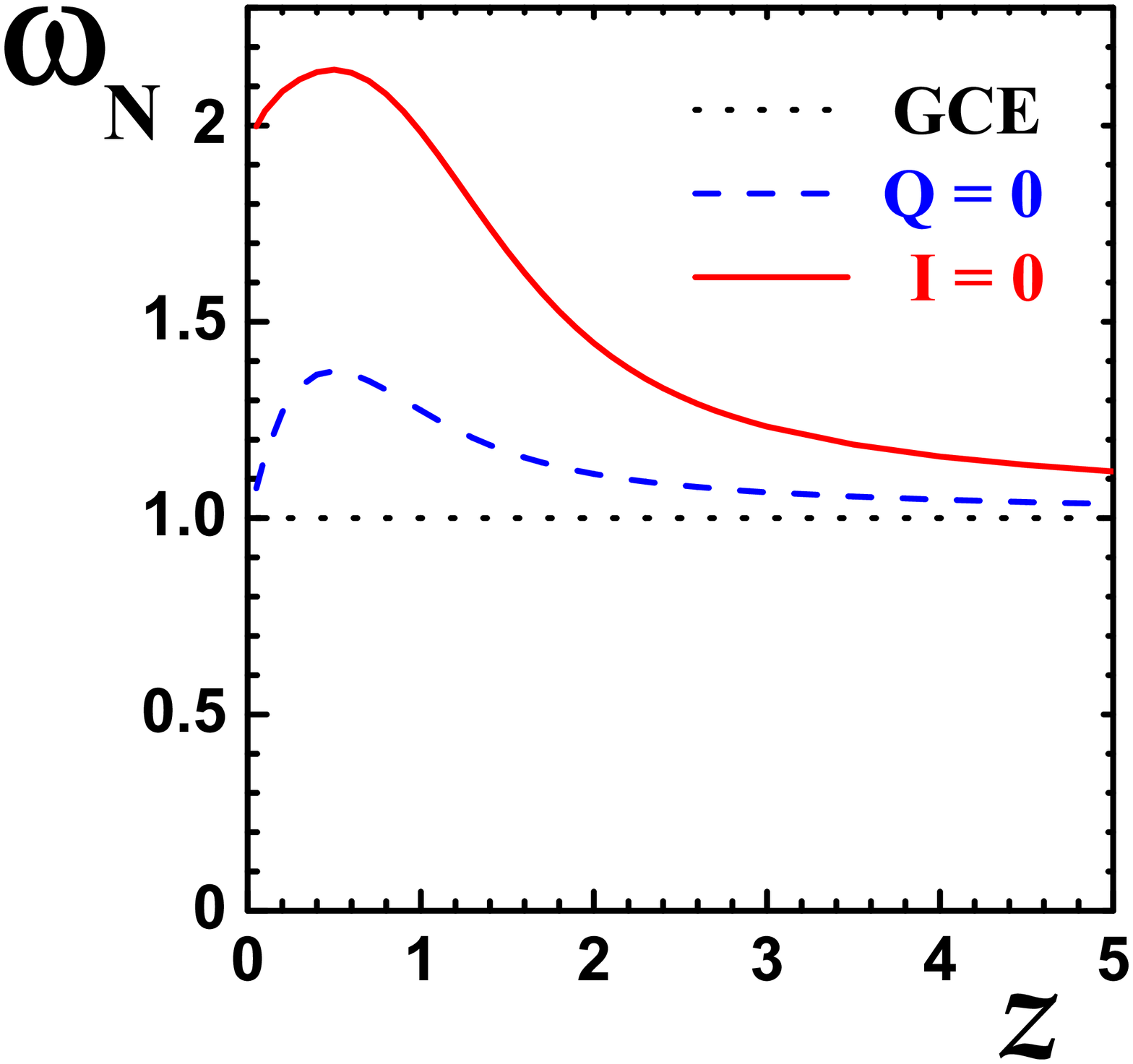,width=0.48\textwidth}
\caption{The ratio $R_N$ ({\it left}) and the scaled variance $\omega_N$ ({\it right})
in the GCE (the dotted lines), CE with Q=0 (the dashed lines), and the statistical ensemble
with $I=0$ (the solid lines)
%
 }
\label{fig-RN}
\end{figure}

The partition functions (\ref{gce}), (\ref{Zce}), and (\ref{ZI=0})
can be simplified by taking $\lambda_j=\lambda$. One finds:
\eq{\label{pfs}
Z_{gce}=\exp(3\lambda z)~,~~~~Z_{Q=0}=\exp(\lambda z)~I_0(2\lambda
z)~,~~~~Z_{I=0}=\exp(\lambda z)~ \left[I_0(2\lambda
z)~-~I_1(2\lambda z)\right]~,
}
where $I_n$ are the modified Bessel functions. The final
expressions for the partition functions correspond to $\lambda=1$.
Taking the derivatives of $Z$ over $T$ and $V$ one finds the
thermodynamical functions of the pion system. The derivatives over
$\lambda$ give the moments of total pion number distribution,
\eq{\label{Ntotk}
\langle N\rangle ~=~\frac{1}{Z}\frac{\partial Z}{\partial
\lambda}\bigg|_{\lambda=1}~, ~~~~~\langle N^2\rangle
~=~\frac{1}{Z}\frac{\partial}{\partial
\lambda}\left(\lambda\frac{\partial Z}{\partial
\lambda}\right)\bigg|_{\lambda=1}~.
}
With Eq.~(\ref{Ntotk}) one calculates the average pion
multiplicity $\langle N\rangle$ and the second moment $\langle
N^2\rangle$  using the partition functions (\ref{pfs}) of
different statistical ensembles:
\eq{\label{N1}
 &\langle N\rangle _{gce}   ~=~3z~,&
 &\langle N^2\rangle _{gce} ~=\;3z\;+\;9z^2~,
\\
 &\langle N\rangle_{Q=0}   ~=~z~+~2z\,\frac{I_1(2z)}{I_0(2z)}~,&
 &\langle N^2\rangle_{Q=0} ~=\;z~+\;5z^2\;+\;4z^2\,\frac{I_1(2z)}{I_0(2z)}
 ~\label{N2},
\\
 &\langle N\rangle _{I=0}   ~=\;z\,\frac{I_1(2z)-I_2(2z)}{I_0(2z)-I_1(2z)}~,&
 &\langle N^2\rangle _{I=0} ~=~  \langle N\rangle _{I=0}
  ~+~ z^2\,\frac{I_0(2z)~-~I_3(2z)}{I_0(2z)-I_1(2z)}~.\label{N3}
}
The ratio $R_N$  and the scaled variance $\omega_N$,
\eq{\label{omega-N}
R_N~\equiv~\frac{ \langle N\rangle}{3z}~,~~~~ \omega_N ~ \equiv~
\frac{\langle N^2\rangle -\langle N\rangle^2}{\langle N\rangle}~,
}
for the total number of pions calculated in different statistical
ensembles are shown in Fig.~\ref{fig-RN}.  In the GCE one obtains
 $R_N(GCE)=\omega_N(GCE)=1$.

Figure~\ref{fig-RN} shows that conservation laws suppress the mean
particle number. For the same volume $V$ and temperature $T$ the
average number of pions in the CE with $Q=0$ is smaller than the
GCE  value $3z$, thus $R_N^Q<1$.  Even stronger suppression
effect is observed in the ensemble with $I=0$, i.e.
$R_N^I<R_N^Q<R_N(GCE)=1$. The opposite effect is seen for the
scaled variance of the pion number fluctuations:
$\omega_N^I>\omega_N^Q>\omega_N(GCE)=1$. To find the asymptotic
behavior of $R_N$ and $\omega_N$ at large $z$ in the statistical
ensembles with $Q=0$ and $I=0$ one can use the expansion of the
modified Bessel functions  at $z\gg1$ \cite{AS}:
\eq{\label{In1}
I_n(2z)~=~\frac{\exp(2z)}{\sqrt{4\pi z}}\left[1~-~\frac{4n^2-1}{16z}~
+~O\left(\frac{1}{z^2}\right)\right]~,
}
in Eqs.~(\ref{N1}-\ref{N3}). This gives, $R_N\rightarrow 1$ and
$\omega_N\rightarrow 1$ in the thermodynamic limit,
$z\rightarrow\infty$, for both $Q=0$ and $I=0$ ensembles. It means
that  the suppression of the pion average multiplicity and
the enhancement of the multiplicity fluctuations are the
finite volume
effects.

The partition functions (\ref{gce}), (\ref{Zce}), and (\ref{ZI=0})
can be presented as,
\eq{\label{pfN}
Z~=~\sum_{N=0}^{\infty}\sum_{N_0=0}^{N}
 F(N,N_0) \cdot \frac{z^N}{N!}~=~\sum_{N=0}^{\infty}g(N) \cdot \frac{z^N}{N!}~.
}
The degeneracy factor $g(N)$ for different statistical ensembles
can be found from Eq.~(\ref{pfs}). To calculate $F(N,N_0)$ one
needs Eqs.~(\ref{gce}, \ref{Zce}, \ref{ZI=0}). The  values of
$g(N)$ and $F(N,N_0)$ in different statistical ensembles are
presented in Table I for $N\le 4$,  see also \cite{Pais}.
Note that $g(N)=3^N$ in the GCE, and $g(0)=1$ in all ensembles.
\begin{table}[h!]
\centering
\begin{tabular}{|c||c|c|c|c|c|c|c||c|c|c|c|c|c||c|c|c|c|c|c||}
  \hhline{|====================:t|}
 &\multicolumn{7}{c||}{GCE} & \multicolumn{6}{c||}{$Q=0$} &
 \multicolumn{6}{c||}{$I=0$}
 \\
 \hhline{~|-------------------}
 \;N\; & \,g(N)\, &  \multicolumn{6}{|c||}{$F(N,N_0)$}
        & \,g(N)\, & \multicolumn{5}{|c||}{$F(N,N_0)$}
        & \,g(N)\, & \multicolumn{5}{|c||}{$F(N,N_0)$}
      \\
  \hhline{|====================|}
   &  & \,$N_0$\, & \multicolumn{5}{|c||}{
   \hspace{0.15cm}0\hspace{0.6cm} 1\hspace{0.6cm} 2\hspace{0.5cm} 3\hspace{0.3cm} 4 }
     &  & \multicolumn{5}{|c||}{
   \hspace{0.05cm}0\hspace{0.4cm} 1\hspace{0.5cm} 2\hspace{0.5cm} 3\hspace{0.35cm} 4 }
     &  & \multicolumn{5}{|c||}{
   \hspace{0.05cm}0\hspace{0.6cm} 1\hspace{0.4cm} 2\hspace{0.4cm} 3\hspace{0.35cm} 4
   }
   \\
 \hhline{|~||~|=|=====||~|=====||~|=====|}
    1 & 3 &   & 2  & 1 & \multicolumn{3}{|c||}{} &
    1 & 0 & 1 & \multicolumn{3}{|c||}{} &
    0 & 0 & 0 & \multicolumn{3}{|c||}{}
   \\
 \hhline{~|~||~|---~~||~|---~~||~|---~~|}
    2 & 9 &   & 4 & 4 & 1 & \multicolumn{2}{|c||}{} &
    3 & 2 & 0 & 1 & \multicolumn{2}{|c||}{} &
    1 & $2/3$ &\,0\, & $1/3$ & \multicolumn{2}{|c||}{}
   \\
 \hhline{~|~||~|----~||~|----~||~|----|}
    3 & 27 &   & 8 & 12 & 6 & 1
  &   & 7  & 0 & 6  & 0 & 1
  &   & 1  & 0 & 1  & 0 & 0 &
   \\
 \hhline{~|~||~|-----||~|-----||~|-----|}
   4 & 81 &        &\, 16 \,&\, 32 \,&\, 24 \,&\, 8 \,&\, 1\;
     & 19 &\, 6  \,&\, 0  \,&\, 12 \,&\, 0 \,&\, 1\;
     & 3  &\, $6/5$ \, & \,\,0\,\, & \,$8/5$\, & \,\,0\,\, &
  \,$1/5$\;
   \\
   \hhline{|====================:b|}
\end{tabular}
\caption{The degeneracy factors $g(N)$ and $F(N,N_0)$ (\ref{pfN})
in the GCE, in the CE with $Q=0$, and in the statistical ensemble with
 $I=0$. }\label{t-1}
\end{table}
Table \ref{t-1} is helpful when one considers very small systems
with only a few pions. For $z\ll1$ one finds from Eq.~(\ref{pfN})
and Table~\ref{t-1}:
\eq{\label{small-z}
Z_{Q=0} \cong 1+1\cdot z+3\cdot \frac{z^2}{2!}
 + 7\cdot\frac{z^3}{3!} + 19\cdot\frac{z^4}{4!}~,~~~~
Z_{I=0}\cong 1+1\cdot\frac{z^2}{2!}+1\cdot\frac{z^3}{3!}
 + 3\cdot\frac{z^4}{4!}~.
%
}
The mean multiplicity and higher moments can be  calculated
as:
%
 \eq{
 \langle N^k\rangle
 \;=\;  \frac{1}{Z}\sum_{N=0}^{\infty} N^k\cdot g(N) \cdot \frac{z^N}{N!}
 ~\cong~\frac{1}{Z}\sum_{N=0}^{4} N^k\cdot g(N) \cdot \frac{z^N}{N!}\;.
 }
This gives:
\eq{\label{N-Q}
 &\langle N\rangle_{Q=0}
\cong~z~+~2z^2~-~z^4~,~~~~ && \langle N^2\rangle_{Q=0}
\cong~z~+~5z^2~+~4z^3~,
\\
& \langle N\rangle_{I=0}
\cong~ z^2 ~+~ \frac{z^3}{2}~, ~~~~
&& \langle N^2\rangle_{I=0}
\cong~  2z^2~+~\frac{3}{2}z^3~+~z^4~.\label{N-I}
}
The results (\ref{N-Q}-\ref{N-I}) can be also obtained from
Eqs.~(\ref{N1}-\ref{N3}) by expanding the  modified Bessel
functions  at  $z\ll1$  \cite{AS},
\eq{\label{In2}
I_n(2z)~=~\frac{z^n}{n!}~+~\frac{z^{n+2}}{(n+1)!}~+~O\left(z^{n+4}\right)~.
}
 For $R_N$ and $\omega_N$
(\ref{omega-N}) one finds from Eqs.~(\ref{N-Q}-\ref{N-I}):
\eq{\label{R-N}
 &R_N^Q~\cong ~\frac{1}{3}~+~\frac{2}{3}~z~,~
 &&\omega_N^Q~\cong ~1~+~2z~-~4z^2~,\\
 & R_N^I~\cong~\frac{1}{3}~z~,~
 &&\omega_N^I~\cong~2~+\frac{z}{2}~-~\frac{z^2}{4}~.
 \label{w-N}
}
The behavior of $R_N$ (\ref{R-N}) and $\omega_N$ (\ref{w-N}) at
small $z$ can be seen in Fig.~\ref{fig-RN}. The scaled variance
$\omega_N$ has a maximum at  $z\approx 0.5$ in both $Q=0$ and
$I=0$ statistical ensembles.
%
\section{fluctuations and correlations of $\pi^0$, $\pi^+$, $\pi^-$}\label{sec-pions}
\begin{figure}[ht!]
\epsfig{file=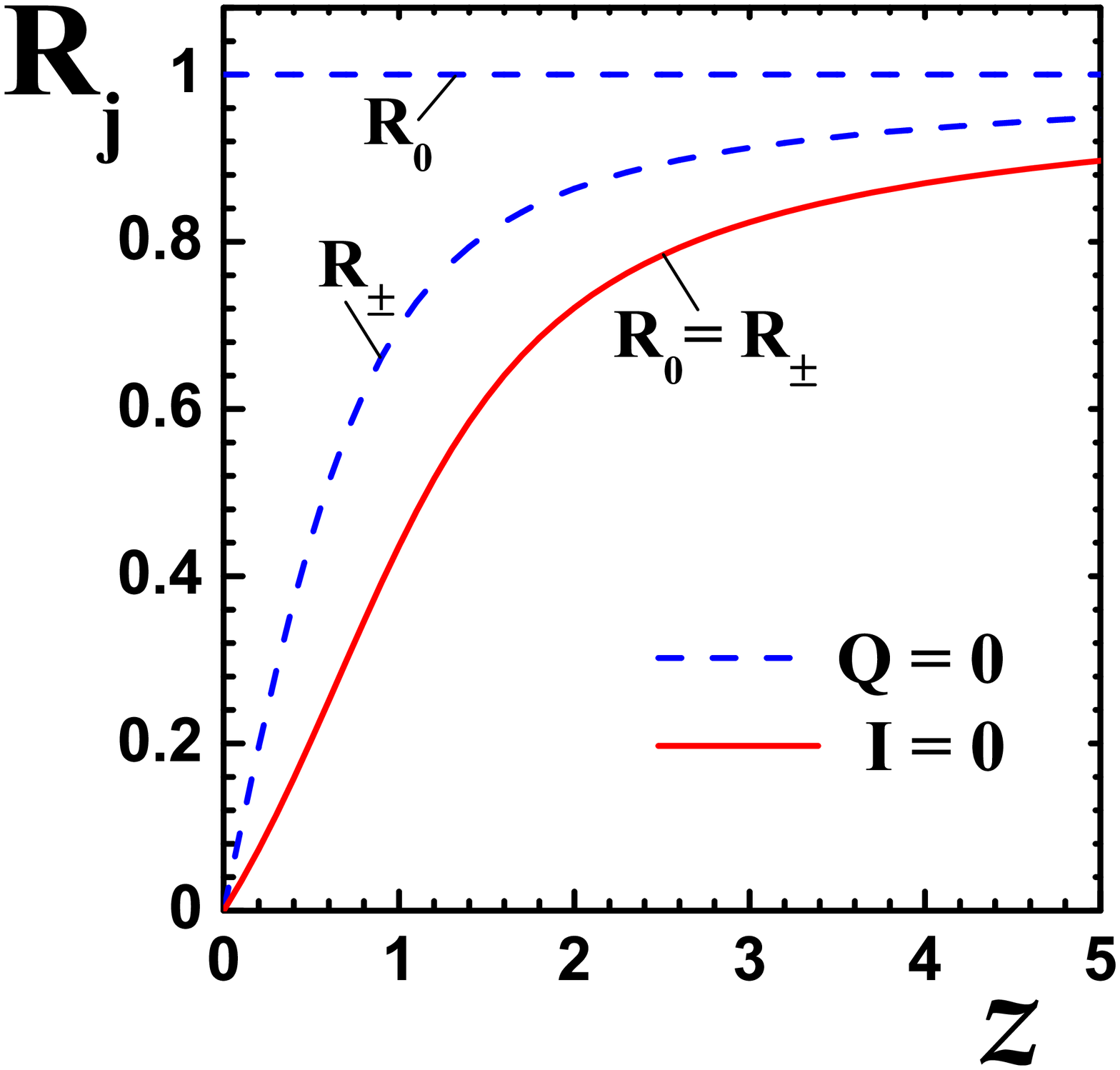,width=0.48\textwidth}~~
\epsfig{file=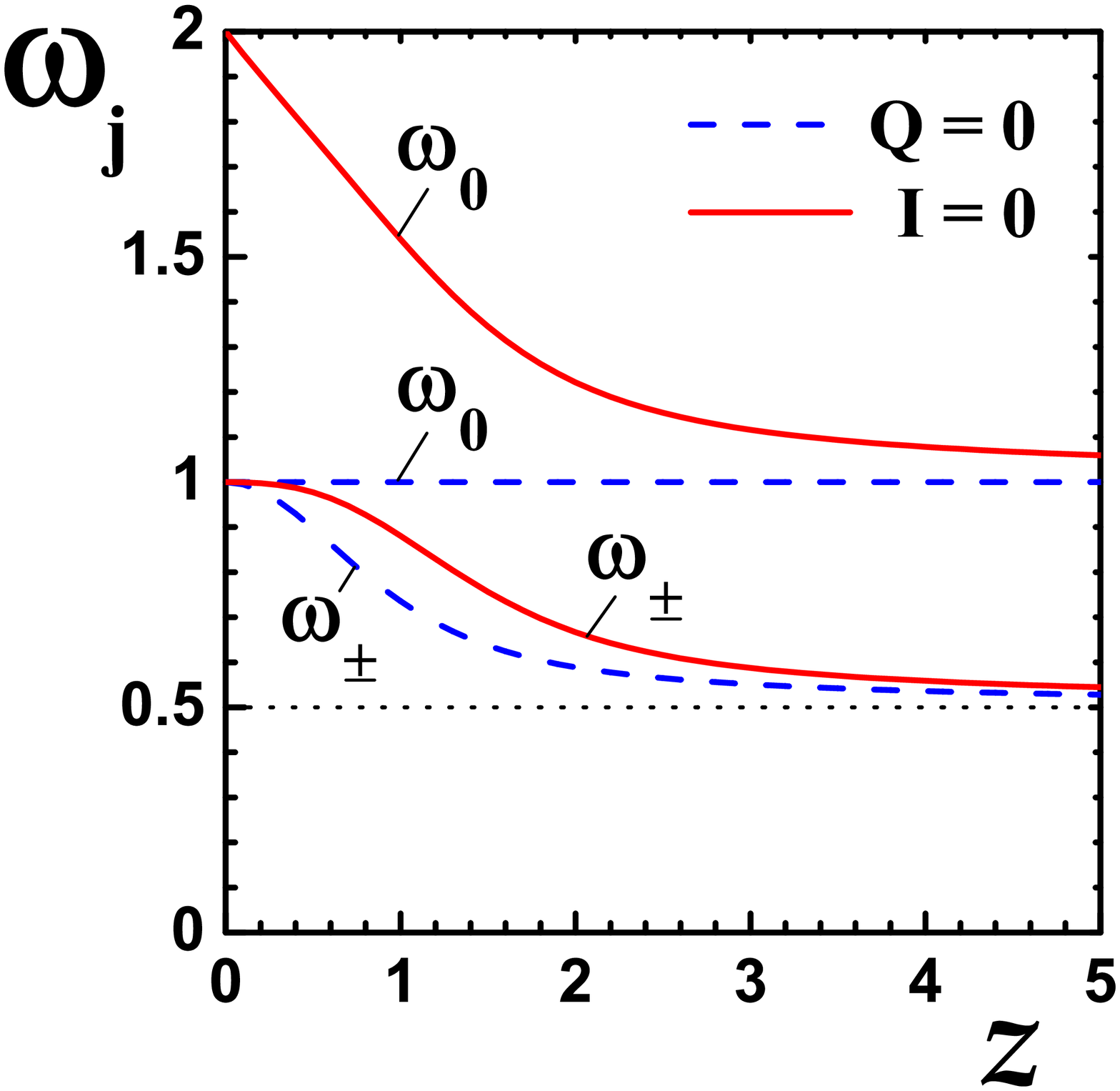,width=0.48\textwidth}
\caption{ The ratios $R_0$ and $R_{\pm}$ ({\it left}) and the scaled variances
$\omega_0$ and $\omega_{\pm}$  ({\it right}) in the CE with $Q=0$ (the dashed
lines) and in the ensemble with $I=0$ (the solid lines).  \label{fig-Rj} }
\end{figure}

%
%
To calculate the mean multiplicities, correlations, and
fluctuations for neutral and charged pions one has to return back
to presentations of the partition functions by Eqs.~(\ref{gce},
\ref{Zce}, \ref{ZI=0}). One finds,
 \eq{
 \langle N_j\rangle
 \;=\; \frac{1}{Z}\,\frac{\partial Z}{\partial
 \lambda_j}\bigg|_{\vec{\lambda}=1}\;,
 ~~~~ \langle N_i\,N_j\rangle
  \;\equiv\; \frac{1}{Z}\,\frac{\partial}{\partial \lambda_i}
              \left(\lambda_j \frac{\partial Z}{\partial \lambda_j}\right)
              \bigg|_{\vec{\lambda}=1}~.
 \label{NiNj}
 }
Using Eq.~(\ref{NiNj}) one obtains for the mean multiplicities of
neutral and charged particles:
 \eq{\label{N0Q}
& \langle N_0\rangle_{gce} \;=\; \langle N_{\pm}\rangle_{gce}
 \;=\;z~,~~~~ \langle N_0\rangle_{Q=0} \;=\;z~, ~~~\; \langle N_{\pm}\rangle_{Q=0}
 \;=\; z\,\frac{I_1(2z)}{I_0(2z)}\;\\
&
 \langle N_0\rangle_{I=0} \;=\; \langle N_{\pm}\rangle_{I=0}
 \;=\; \frac{z}{3}\,\frac{I_1(2z)-I_2(2z)}{I_0(2z)-I_1(2z)}\;, \label{N0I}
 }
where $\langle N_{\pm}\rangle = \langle N_+\rangle=\langle
N_-\rangle$.
The ratios
\eq{\label{Rj}
R_0~\equiv ~\frac{\langle
N_0\rangle}{z}~,~~~~~ R_{\pm}~=~\frac{\langle N_{\pm}\rangle}{z}
}
are shown in Fig.~\ref{fig-Rj}  ({\it left}) for $Q=0$ and $I=0$
statistical ensembles.

Note that $R_0=R_{\pm}=1$ in the GCE. In the CE,  $R_0^Q=1$, i.e.
the mean number of neutral pions is not affected by the charge
conservation law.  The mean number of charged pions is suppressed
in the CE, $R_{\pm}^Q<1$ \cite{RD}. The behavior in the
statistical ensemble with $I=0$  differs from that in the CE with
$Q=0$. The pion mean numbers are the same for all charge pion
states $\pi^0$, $\pi^+$, and $\pi^-$.  The suppression of these pion
multiplicities at $I=0$ is stronger than that  for $\pi^{-}$ or $\pi^+$
in the CE with $Q=0$:
\eq{\label{R0R+}
R_0^I~=~R_{\pm}^I~<~R_{\pm}^Q~<~R_0^Q~=1~. }
The asymptotic behavior of (\ref{N0Q}, \ref{N0I}) at large $z$ can
be found using Eq.~(\ref{In1}). One obtains  $R_j\rightarrow 1$ at
$z\rightarrow\infty$ for all $j=0,+,-$ in both $Q=0$ and $I=0$
statistical ensembles. Thus, the suppression of the mean numbers
of charged and neutral pions  is the  finite volume
effect.
At small $z$ one can use Eq.~(\ref{In2}), this gives at $z\ll1$:
\eq{\label{Rj-small-z}
R_{\pm}^Q~\cong~z~,~~~~R_0^I~=~R_{\pm}^I~\cong~\frac{z}{3}~.
}
%

The second derivatives in Eq.~(\ref{NiNj}) can be
calculated using the partition functions (\ref{gce}, \ref{Zce},
\ref{ZI=0}) for different statistical ensembles. They give the
second moments and correlations in the GCE:
\eq{\label{N02gce}
\langle N_0^2\rangle_{gce}~=~\langle N_{\pm}^2\rangle_{gce}
 \;=\; z \;+\; z^2\;,~~~~\langle N_0 N_{\pm}\rangle_{gce}~=~z^2~,
}
in the CE:
\eq{\label{N02ce}
 \langle N_0^2\rangle_{Q=0}
 \;=\; z \;+\; z^2\;,
 ~~~~\langle N_{\pm}^2\rangle_{Q=0}
 ~=~z^2\;,~~~~
 \langle N_0\, N_{\pm}\rangle_{Q=0}
 \;=\;
 z~\langle N_{\pm}\rangle_{Q=0}\;,
 }
and in the pion system with $I=0$:
 \eq{\label{N02}
& \langle N_0^2\rangle_{I=0}
 \;=\; \langle N_0\rangle_{I=0}
  \;+\; \frac{1}{15}\,\frac{z\left(7z\,+\,4\right)\,I_0(2z)
  \;-\; \left(7z^2\,+\,2z\,+\,4\right)\,I_1(2z)}
         {I_0(2z)\;-\;I_1(2z)}\;,
 \\
 &\langle N_{\pm}^2\rangle_{I=0}
 \;=\; \langle N_{\pm}\rangle_{I=0}
  \;+\; \frac{z^2}{5}\,\frac{I_2(2z)\;-\;I_3(2z)}{I_0(2z)\;-\;I_1(2z)}\;,
  \label{Npm2}
\\
&
\langle N_0\,N_{\pm}\rangle_{I=0}
 \;=\; \frac{z}{15}\,
 \frac{z\,I_1(2z)\;-\;(z-3)\,I_2(2z)}{I_0(2z)\;-\;I_1(2z)}\;,
 \label{N0pm}
}
where $\langle N_{\pm}^2\rangle=\langle N_+^2\rangle=\langle
N_-^2\rangle=\langle N_+N_-\rangle$.
%

%

Using the above expressions  one can easily construct the scaled
variances $\omega_j$  and correlation coefficients $\rho_{ij}$,
 \eq{\label{omega-j}
 \omega_j
 \;\equiv\; \frac{\langle N_j^2\rangle \;-\; \langle N_j\rangle^2}
        {\langle N_j\rangle}\;,\qquad
 \rho_{ij}
 \;\equiv\; \frac{\langle N_i N_j\rangle \;-\; \langle N_i\rangle \langle N_j\rangle}
       {\sqrt{\omega_i\,\omega_j\,\langle N_i\rangle\, \langle N_j\rangle}}\;,
 }
that define the main characteristics of the pion multiplicity
distributions.

From Eqs.~(\ref{N0Q}-\ref{N0I}, \ref{N02}-\ref{Npm2}) and
asymptotic expansion of the modified Bessel functions (\ref{In1})
one finds the behavior of the scaled variances $\omega_j$
(\ref{omega-j}) in the thermodynamic limit $z\rightarrow \infty$.
As seen from Fig.~\ref{fig-Rj}, the isospin conservation with
$I=0$ gives the same pion number fluctuations as the CE with
$Q=0$, $\omega_0^I\rightarrow 1$ and $\omega_{\pm}^I\rightarrow
1/2$, at $z\rightarrow\infty$.

However, the results for finite systems are rather different. In
the CE with $Q=0$ the fluctuations of neutral particles are the
same as in the GCE, $\omega_0^Q=1$. The scaled variance
$\omega_{\pm}^Q$ in the CE with $Q=0$ was calculated in
Ref.~\cite{CE}. At $z\ll1$ using Eq.~(\ref{In2}), one finds:
%
%
$\omega_{\pm}^Q~\cong~1~-~z^2/2~$.
%
%

In the statistical ensemble with $I=0$ the scaled variances
$\omega_0$ and $\omega_{\pm}$  can be calculated for small system
with $z<1$ using  Eqs.~(\ref{N0Q}-\ref{N0I}, \ref{N02}-\ref{Npm2})
and asymptotic expansion of the modified Bessel functions
(\ref{In2}).
%
%
%
%
%
However, it is more instructive to obtain approximate expressions
for the distributions of $\pi^0$ and $\pi^{\pm}$ numbers using
Eq.~(\ref{pfN}) and $F(N,N_0)$ values from Table \ref{t-1}. Let us
consider the pion states with total number of pions $N\le 4$, i.e.
we neglect small terms of the order of $z^5$. The probability
distribution $P_0(N_0)$
then reads:
\eq{
 P_0(N_0)~\cong~\frac{1}{Z}\sum_{N=0}^4\;F(N,N_0)\;\frac{z^N}{N!}~.
\label{PN0}
}
For the statistical ensemble with $I=0$ one finds from
Eq.~(\ref{PN0}) and Table \ref{t-1},
\eq{\label{PN0a}
 P_0(1) &~\cong~\frac{1}{Z_{I=0}}~1\cdot\frac{z^3}{3!}~\cong~ \frac{z^3}{6}~,
 \\
 P_0(2) &~\cong~\frac{1}{Z_{I=0}}~\left[\frac{1}{3}\cdot
 \frac{z^2}{2!}~+~\frac{8}{5}\cdot\frac{z^4}{4!}\right] \cong~\frac{z^2}{6}~-~\frac{z^4}{60}~,
 \\
 P_0(3) &~\cong~0~,~~~~~~~~P_0(4)~\cong~\frac{1}{Z_{I=0}}~\frac{1}{5}\cdot\frac{z^4}{4!}
~\cong \frac{z^4}{120}~ \label{PN01a}.
}
Similar expressions for $P_{\pm}(N_{\pm})$ in the statistical
ensemble with $I=0$ are equal to:
\eq{\label{PNpm-I0}
%
 P_{\pm}(1)
 & ~\cong~ \frac{1}{Z_{I=0}}~\left[\frac{2}{3}\cdot\frac{z^2}{2!}
   ~+~ 1\cdot\frac{z^3}{3!} ~+~ \frac{8}{5}\cdot\frac{z^4}{4!} \right]
   ~\cong~ \frac{z^2}{3}~+~\frac{z^3}{6}~-~\frac{z^4}{10}~,
    \\
 P_{\pm}(2)
 & ~\cong~ \frac{1}{Z_{I=0}}~\left[\frac{6}{5}\cdot\frac{z^4}{4!}\right]
   ~\cong~ \frac{z^4}{20}~,~~~~
  P_{\pm}(3)  ~\cong~ P_{\pm}(4)~\cong~0~. \label{PNpm}
%
}
The distributions $P_0$ and $P_{\pm}$ are rather different.
The one and two particle states have evidently
different probabilities for neutral and   (negative) positive pions.
At $z\ll1$ the main configurations consist of 1 (negative) positive
pion and 2 neutral pions. These states correspond to
the total number of pions $N=2$, and their probabilities are
proportional to $z^2$. The states with only 1 neutral pion have
smaller probability, $z^3$, as they can only appear for $N\ge 3$.
Inspite of differnces in the $P_0$ and $P_{\pm}$
distributions they give, however, the same average numbers of
neutral and  (negative) positive pions:
\eq{\label{N0av}
&\langle N_0\rangle_{I=0}\cong1\cdot P_0(1)~+~2\cdot
P_0(2)~+~4\cdot P_0(4)~\cong~\frac{z^2}{3}~+~\frac{z^3}{6}~,\\
& \langle N_{\pm}\rangle_{I=0}\cong1\cdot P_{\pm}(1)~+~2\cdot
P_{\pm}(2) \cong~\frac{z^2}{3}~+~\frac{z^3}{6}~. \label{Npmav}
}
Equations  (\ref{N0av},\ref{Npmav})  clearly demonstrate that the
same average numbers of $\pi^0$ and $\pi^{\pm}$ come from
different pion states.
The results for higher moments of $P_0$ and $P_{\pm}$
distributions are  different:
\eq{\label{Nk0}
& \langle N_0^2\rangle_{I=0}
 \cong 1^2\cdot P_0(1)~+~2^2\cdot P_0(2)~+~4^2\cdot P_0(4)\cong~\frac{2z^2}{3}
  ~+~\frac{z^3}{6}~+~\frac{z^4}{15}~,
  \\
& \langle N_{\pm}^2\rangle_{I=0}\cong1^2\cdot
P_{\pm}(1)~+~2^2\cdot
P_{\pm}(2)~\cong~\frac{z^2}{3}~+~\frac{z^3}{6}~+~\frac{z^4}{10}~.
\label{Nkpm}
}
From Eqs.~(\ref{N0av}-\ref{Nkpm}) one  finds:
\eq{\label{omega-j-small}
 \omega_0^I ~\cong~2~-~\frac{z}{2}~,~~~~
 \omega_{\pm}^I~\cong~1~-~\frac{z^2}{30}~.
}
%
%

\begin{figure}[ht!]
\epsfig{file=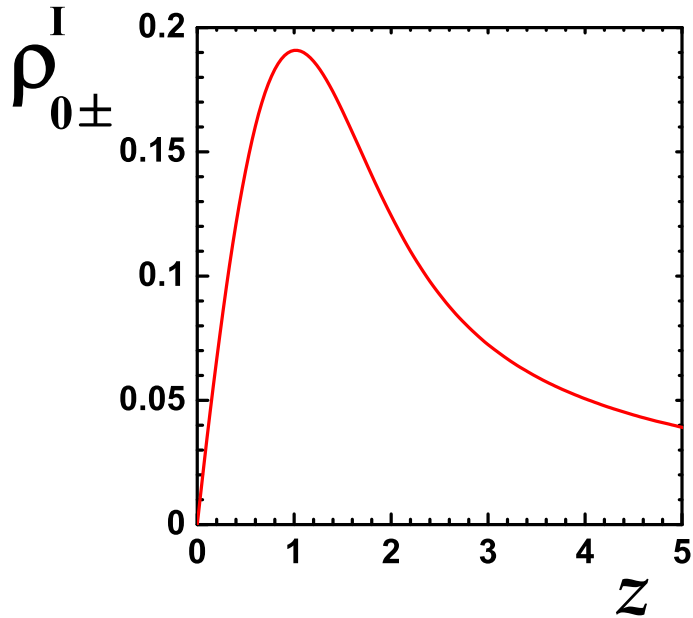,width=0.55\textwidth}
\caption{The correlation coefficient $\rho^I_{0\,\pm}$
(\ref{omega-j}) between the neutral and charged pions in the
statistical ensemble with $I=0$. } \label{fig-rho}
\end{figure}

There are no correlations between the numbers of $\pi^0$, $\pi^+$,
$\pi^-$ in the GCE. All correlation coefficients defined by
Eq.~(\ref{omega-j}) are equal to zero,
$\rho_{0+}=\rho_{0-}=\rho_{+-}=0$. The charge is exactly conserved
in the $Q=0$ and $I=0$ statistical ensembles. This brings the
strongest correlations between the numbers of $\pi^+$ and $\pi^-$,
i.e. $\rho_{+-}=1$, and this means equal numbers $N_+$ and $N_-$ in
each microscopic state of the system. The correlations between the
numbers of $\pi^0$ and $\pi^{\pm}$, are absent in the CE with
$Q=0$, but exist in the statistical ensemble with $I=0$. The
correlation coefficient $\rho^I_{0\pm}$ at $I=0$ is presented in
Fig.~\ref{fig-rho}. It has the maximal value $\rho^I_{0\pm}\approx
0.19$ at $z\approx 1$ and goes to zero at $z\rightarrow\infty$.

The role of the pion number correlations can be illustrated
returning to the scaled variance $\omega_N$ presented in
Fig.~\ref{fig-RN}, {\it right}. Taking into account the relation
$N\equiv N_0+N_++N_-$ one finds:
\eq{\label{omega-N-all}
\omega_N~=~\omega_0~\frac{\langle N_0\rangle}{\langle N
\rangle}~+~ 2\omega_{\pm}~\frac{\langle N_{\pm}\rangle}{\langle N
\rangle}~+~ 2\rho_{+-}~\omega_{\pm}~\frac{\langle
N_{\pm}\rangle}{\langle
N\rangle}~+~2\rho_{0\pm}~\sqrt{\omega_0\omega_{\pm}}~\frac{\langle
N_{\pm}\rangle}{\langle N\rangle}~.
}
The GCE corresponds to $\rho_{+-}=\rho_{0\pm}=0$ in
Eq.~(\ref{omega-N-all}). Besides, the multiplicities $\langle
N_j\rangle =\langle N\rangle/3$ and scaled variances $\omega_j=1$
are the same for $j=0,+,-$. This leads to $\omega_N=1$ in the GCE.

The CE with $Q=0$ corresponds to $\omega^Q_0=1$, $\rho^Q_{+-}=1$ and
$\rho^Q_{0\pm}=0$ in Eq.~(\ref{omega-N-all}). This gives,
\eq{\label{omega-N-Q}
\omega_N^Q~=~\frac{\langle N_0\rangle_{Q=0}}{\langle N
\rangle_{Q=0}}~+~ 4\omega_{\pm}^Q~\frac{\langle
N_{\pm}\rangle_{Q=0}}{\langle N \rangle_{Q=0}}
 ~\cong~ (1-2z+4z^2)~+~4(z-2z^2)~=~1+2z-4z^2~.
}
The  term $\langle N_0\rangle_{Q=0}/\langle N
\rangle_{Q=0}$ in  Eq.~(\ref{omega-N-Q})
decreases at small $z$.  The maximum of $\omega_N^Q$ appears due
to $\rho^Q_{+-}=1$, note the factor 4 in the second term in r.h.s.
of Eq.~(\ref{omega-N-Q}). Due to the exact charge conservation,
negative and positive pions may appear only as $\pi^+\pi^+$-pairs.
For $N_{ch}\equiv N_++N_-$, the scaled variance
\eq{\label{omega-ch}
 \omega_{ch}^Q~=~\frac{\langle N_{ch}^2\rangle_{Q=0}~-~\langle
 N_{ch}\rangle^2_{Q=0}}{\langle
 N_{ch}\rangle_{Q=0}}~=~2~\omega^Q_{-}~=~2~\omega^Q_+
 }
is then two times larger than that for positive (negative) pions.
The same relation, $\omega_{ch}=2\omega_{\pm}$, is also valid at
$I=0$.

For $I=0$ one obtains from Eq.~(\ref{omega-N-all}):
\eq{\label{omega-N-I}
&\omega_N^I~=~\frac{1}{3}\left(\omega^I_0~+~ 4\omega_{\pm}^I
 ~+~ 4\rho^I_{0\pm}~\sqrt{\omega^I_0\omega^I_{\pm}}\right) \nonumber \\
& ~\cong~
\left(\frac{2}{3}-\frac{z}{6}+\frac{7z^2}{180}\right)~+~\left(\frac{4}{3}-\frac{2z^2}{45}\right)~+~
\left(\frac{2z}{3}-\frac{11z^2}{45}\right)~=~
2~+~\frac{z}{2}~-~\frac{z^2}{4} ~.
}
where the relation $\langle N_0\rangle_{I=0}=\langle
N_{\pm}\rangle_{I=0}=\langle N\rangle_{I=0}/3$ is used. The sum of
$\omega^I_0/3$ and $4\omega^I_{\pm}/3$  decreases at small $z$. An
increase of $\omega^I_N$ and its maximum at small $z$ comes due to
the third term in the r.h.s. of Eq.~(\ref{omega-N-I}), i.e. due to
the correlations between the neutral and charged pions,
$\rho_{0\pm}^I>0$. Note that a conclusive comparison with
experimental data for small pion systems would require an
inclusion of additional conservation laws like 4-momentum, angular
momentum, etc. \cite{Becattini}.

\section{Bose Statistic}\label{sec-Bose}
\begin{figure}[ht!]
\epsfig{file=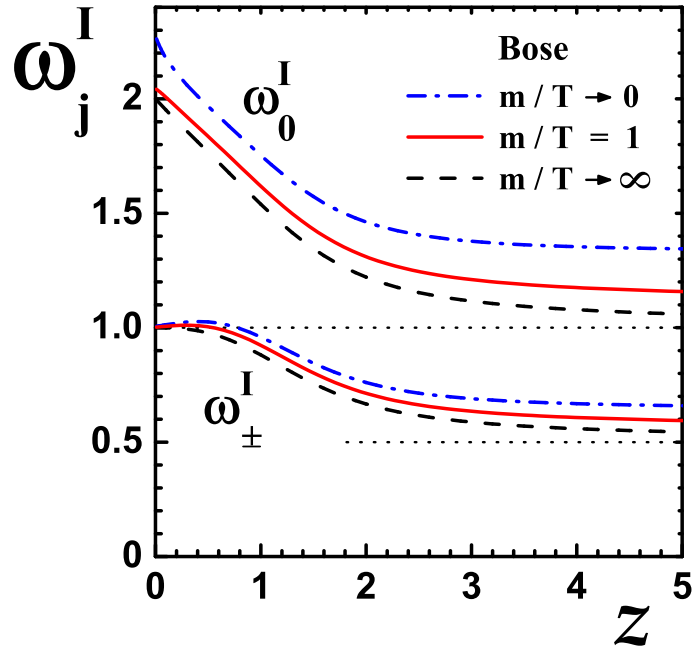,width=0.55\textwidth}
\caption{The scaled variances $\omega^I_0$ and $\omega^I_{\pm}$ in
the Bose gas with $I=0$.  The solid lines correspond to $m/T=1$,
dashed-dotted lines to $m/T\rightarrow 0$, and dashed lines to
$m/T\rightarrow \infty$. } \label{fig-Bose}
\end{figure}

In this section we discuss the role of Bose effects in the
statistical ensemble with fixed isopsin. The role of Bose effects
for the fluctuations in the CE was considered in
Ref.~\cite{CE-Bose}. A generalization of Eq.~(\ref{Z-Boltz}) for
Bose statistics gives the following expression:
 \eq{\label{Z-Bose}
& Z^{Bose}_{I=0}~ =~
 \frac{1}{8\pi^2}\int_0^{2\pi}d\alpha\int_0^{2\pi}d\gamma\int_0^{\pi}d\beta
    \sin\beta
 \\
 &\times\;
    \exp\left[\sum_{n=1}^{\infty}\frac{z_n}{n}\left(\lambda_0^n \cos^n\beta
    \;+\; \left(\frac{1+\cos\beta}{2}\right)^n
    \left(\lambda^n_+\exp[in(\alpha+\gamma)]
    \;+\; \lambda^n_-\exp[-in(\alpha+\gamma)]\right)\right)\right]\;,
 \nonumber
 }
where
\eq{\label{zn}
z_n~=~\frac{V}{2\pi^2}~\frac{Tm^2}{n}~K_2\left(\frac{nm}{T}\right)~.
}
The Boltzmann approximation corresponds to the first term $n=1$
in the sum in Eq.~(\ref{Z-Bose}).  The results for
$\omega_j^I$ are presented in
Fig.~\ref{fig-Bose}.

In the case of quantum statistics, the partition function depends
not only on the one particle partition  function $z$ (\ref{z}),
but additionally on the value of $m/T$. In Fig.~\ref{fig-Bose} the
results for the scaled variances $\omega_j^I$ are shown for a
typical value $m/T=1$ and for two limiting cases: $m/T\rightarrow
0$ when the Bose effects are the strongest ones, and
$m/T\rightarrow \infty$ when the Bose effects disappear and
results coincide with the Boltzmann approximation presented in
Fig.~\ref{fig-Rj}, {\it right}. As seen from Fig.~\ref{fig-Bose},
Bose statistics makes the pion number fluctuations larger. These
effects are always stronger for smaller values of the $m/T$ ratio.
For $m/T\rightarrow 0$ we find $\omega^I_0\cong 2.26$ at
$z\rightarrow 0$, and $\omega_0^I\cong 1.368$ at $z\rightarrow
\infty$. The corresponding results for the charged particles are
the following: $\omega^I_{\pm}=1$ at $z\rightarrow 0$, and
$\omega_{\pm}^I\cong 0.684$ at $z\rightarrow \infty$. The results
for $z\rightarrow\infty$ coincide with those in the canonical
ensemble with $Q=0$.

The Bose statistics effects become visible when more than one
identical pion (i.e. pions in the same charge states) can appear
in the same quantum state. Note a difference between the neutral
and charged pions for very small  systems with $I=0$. At $z<<1$
the main configurations consist of 1 (negative) positive pion and
2 neutral pions. These states correspond to the total number of
pions $N=2$ and their probabilities are proportional to $z^2$. The
states with only 1 neutral pion have smaller probability, $z^3$,
as they can only appear for $N\ge 3$. Because of this difference
one observes no Bose effects for $\omega_{\pm}^I$ at $z\rightarrow
0$ in Fig.~\ref{fig-Bose}, whereas these effects are seen for
$\omega_0^I$.

%
\section{Summary}\label{sec-summary}
Particle number fluctuations and correlations in the statistical
system of pions with zero total isospin $I=0$ have been studied in
the present paper. For finite systems one observes a suppression
of the average total pion number and  an increase of the
pion number fluctuations.
%
The suppression effects due to isospin conservation are the same
for average numbers of $\pi^0$, $\pi^+$ and $\pi^-$. However, we
find quite different behavior of the  corresponding scaled
variances. For neutral pions there is the enhancement of the
fluctuations,
whereas for charged pions the
isospin conservation suppresses  fluctuations
similar to that in the canonical ensemble with $Q=0$.  The positive
correlations between the numbers of neutral and (negative) positive pions are
observed for $I=0$.  This effects is absent in the canonical ensemble with $Q=0$.
The correlations between the numbers of neutral and charged pions
are responsible for a maximum
of $\omega_N$  for small systems with $I=0$.
In the thermodynamic limit  the correlation coefficient
$\rho^I_{0\pm}$  between neutral and charged pions goes to zero.
This leads  to $\omega_0^I\rightarrow 1$,
$\omega_{\pm}^I\rightarrow 1/2$ for Boltzmann statistics and to
$\omega_0^I\rightarrow 1.368$, $\omega_{\pm}^I\rightarrow
\omega_0^I/2=0.684$ for $m=0$ and Bose statistics.

Finally, we speculate on possible consequences of our results for the
quark-gluon gas with SU(3)-color group. A requirement of colorless
of the system of quarks and gluons is similar to that of the
isospin singlet $I=0$. This requirement cause the suppression
effects for the average yields of quarks and gluons, the same for
different colors. These suppression effects for the yields disappear in the
thermodynamic limit. However, a behavior of the fluctuations are
different. To make situation fully transparent let us
consider a toy model of the gluon gas with SU(2)-color. The
colorless system of gluons will be then in  one-to-one
correspondence with the iso-singlet $I=0$ pion gas. We thus
immediately conclude that  fluctuations of the number of gluons with different
colors are different. Moreover, this difference survives in
the thermodynamic limit too.

\vspace{0.3cm} {\bf Acknowledgments.} We would like to thank
M.~Ga\'zdzicki, W.~Greiner, and I.A.~Pshenichnov for fruitful
discussions. V.V.~Begun thanks the Alexander von Humboldt
Foundation for the support. This work was in part supported by the
Program of Fundamental Research of the Department of Physics and
Astronomy of NAS, Ukraine.

%

%


\begin{thebibliography}{10}



\bibitem{Bethe}
  H.~A.~Bethe,
  Phys.\ Rev.\  {\bf 50}, 332 (1936).


\bibitem{Pais}
  A.~Pais,
  Annals Phys.\  {\bf 9}, 548 (1960).

\bibitem{Kretz}
  M.~Kretzschmar,
  Ann.\ Rev.\ Nucl.\ Part.\ Sci.\  {\bf 11}, 1 (1961).



\bibitem{Turko}
K. Redlich and L. Turko, Z. Phys. C {\bf 5}, 541 (1980);
%
  L.~Turko,
  Phys.\ Lett.\  B {\bf 104}, 153 (1981).
%

\bibitem{Muller}
  B.~Muller and J.~Rafelski,
  Phys.\ Lett.\  B {\bf 116}, 274 (1982).
%

\bibitem{Blumel}
  W.~Blumel, P.~Koch and U.~W.~Heinz,
  Z.\ Phys.\  C {\bf 63}, 637 (1994);
%
  W.~Blumel and U.~W.~Heinz,
  Z.\ Phys.\  C {\bf 67}, 281 (1995)
  [arXiv:hep-ph/9409343].

\bibitem{Elze}
H.~T.~Elze, W.~Greiner, and J. Rafelsky, Phys. Lett. B {\bf 124},
515 (1983);
  H.~T.~Elze and W.~Greiner,
  Phys.\ Rev.\  A {\bf 33}, 1879 (1986).


\bibitem{GM}
  M.~I.~Gorenstein, O.~A.~Mogilevsky, V.~K.~Petrov and G.~M.~Zinovjev,
  Z.\ Phys.\  C {\bf 18}, 13 (1983);
  M.~I.~Gorenstein, S.~I.~Lipskikh, V.~K.~Petrov and G.~M.~Zinovjev,
  Phys.\ Lett.\  B {\bf 123}, 437 (1983).
%

\bibitem{CE}
  V.~V.~Begun, M.~Gazdzicki, M.~I.~Gorenstein and O.~S.~Zozulya,
  Phys.\ Rev.\  C {\bf 70}, 034901 (2004)
  [arXiv:nucl-th/0404056].

\bibitem{RD}
J. Rafelski and M. Danos, Phys. Lett. B {\bf 97}, 279 (1980).

\bibitem{Wigner}
E.P. Wigner, {\it Group Theory and its Application to the Quantum
Mechanics of Atomic Spectra} (Academic Press, New York and London,
1959).


%
\bibitem{Turko2}
  L.~Turko and J.~Rafelski,
  Eur.\ Phys.\ J.\  C {\bf 18}, 587 (2001)
  [arXiv:hep-th/0003079];
  L.~Turko,
  Acta Phys.\ Polon.\  B {\bf 33}, 1533 (2002)
  [arXiv:hep-th/0204255].

\bibitem{AS}
M. Abramowitz and I.E. Stegun, {\it Handbook of Mathematical
Functions} (Dover, New york, 1964).

\bibitem{CE-Bose}
V.V. Begun and M.I. Gorenstein, Phys. Rev. C {\bf 73} 054904
(2006) [arXiv:nucl-th/0510022].

\bibitem{Becattini}
  L.~Ferroni and F.~Becattini,
  PoS  {\bf CPOD2006}, 041 (2006).




\end{thebibliography}
\end{document}